\RequirePackage{fix-cm}
\documentclass[smallextended]{svjour3}       
\smartqed  
\usepackage{cite}
\usepackage{upgreek}	
\usepackage{fixltx2e}	
\usepackage{color,soul} 
\usepackage{textgreek}	
\usepackage{graphicx}
\usepackage{epstopdf}	
\usepackage{mathptmx}   
\usepackage[utf8x]{inputenc}  
\usepackage[T1]{fontenc}
\usepackage{todonotes}	
\usepackage{amsmath}	
\begin{document}
\title{Influence of field effects on the performance of InGaAs-based terahertz radiation detectors
\thanks{This work was supported by the Research Council of Lithuania (project LAT 04/2016).
}}

\author{Linas~Minkevičius\and Vincas~Tamošiūnas \and Martynas~Kojelis \and Ernestas~Žąsinas	\and Virginijus~Bukauskas \and Arūnas~Šetkus \and Renata~Butkutė \and Irmantas~Kašalynas \and Gintaras~Valušis 
}


\institute{L. Minkevičius, V. Tamošiūnas, R.Butkutė, I. Kašalynas, G. Valušis \at
Center for Physical Sciences and Technology, Saulėtekio av. 3, LT-10257 Vilnius, Lithuania\\
Vilnius University, Faculty of Physics, Saulėtekio av. 9, Bld. III, LT- 10222 Vilnius, Lithuania\\
   Tel.: +370-5-2312418\\
   Fax: +370-5-2627123\\
\email{linas.minkevicius@ftmc.lt}           
\and
 	M. Kojelis, E. Žąsinas \at
Vilnius University, Institute of Applied Research, Saulėtekio av. 9, Bld. III, LT- 10222 Vilnius, Lithuania
             \and            
V. Bukauskas, A. Šetkus \\
Center for Physical Sciences and Technology, Saulėtekio av. 3, LT-10257 Vilnius, Lithuania
}

\date{Received: date / Accepted: date}

\maketitle
\begin{abstract}
A detailed electrical characterization of high-performance bow-tie InGaAs-based terahertz detectors is presented along with simulation results. The local surface potential and tunnelling current were scanned over the surfaces of the detectors by means of Kelvin probe force microscopy (KPFM) and scanning tunnelling microscopy (STM), which also enabled the determination of the Fermi level. Current-voltage curves were measured and modelled using the Synopsys Sentaurus TCAD
package to gain deeper insight into the processes involved in detector operation. In addition, we performed finite-difference time-domain (FDTD) simulations to reveal features related to changes in the electric field due to the metal detector contacts. The investigation revealed that field-effect-induced conductivity modulation is a possible mechanism contributing to the high sensitivity of the studied detectors.
\end{abstract}

\section{Introduction}
\label{intro}
The search for highly sensitive detectors operating at room temperature remains one of the top priorities in the development of solid-state-based THz devices. Multiple operating concepts have been tested, which are used in, for example, Schottky detectors \cite{yasui2006detection}, microbolometer arrays \cite{microbolometerKasalynas2016}, semiconductor field-effect transistors (FETs) with nanometric gates \cite{KnapPlasma, ShurPlasma, BoppelIEEE2012, LisauskasRational}, and bow tie (BT)-shaped semiconductor diodes \cite{SeliutaELL2004,SeliutaELL2006,KasalynasELL2009broken}. Recently, it has been demonstrated that InGaAs-based BT diodes can reach 10~V/W in sensitivity with a noise equivalent power (NEP) of 4~nW/(Hz)\textsuperscript{1/2} in direct mode \cite{KasalynasSensors2013} and 230~fW/Hz in heterodyne mode \cite{LinasAPL}. Considering their reliability in typical implementation environments, devices of this type remain among the most promising candidates for room-temperature THz measurements.

It has been shown that the performance of BT diodes relies on non-uniform carrier heating induced by THz radiation, which is concentrated in the semiconducting part of the diode because of the metallic contacts \cite{SeliutaELL2006}. 

In this paper, we show that in addition to carrier heating effects, the self-mixing in the low-conductivity InGaAs layer can play a significant role in useful signal formation. The results of an experimental investigation and numerical simulations of the electrical properties of InGaAs-based THz BT detectors are presented. Kelvin probe force microscopy (KPFM) \cite{melitz2011kelvin} and scanning tunnelling microscopy (STM) were applied to measure the local surface potentials and tunnelling currents on the surfaces of the detectors. The obtained results enabled an evaluation of the position of the Fermi level, thus providing the starting parameters for further theoretical analyses using the Synopsys Sentaurus TCAD package for semiconductor devices. Finite-difference time-domain (FDTD) simulations of the electric field enhancement and antenna-related effects allowed us to highlight the physical effects contributing to the origin of the detector response.

\section{Samples}

\begin{figure}
\centering
\includegraphics[width=1.0\textwidth]{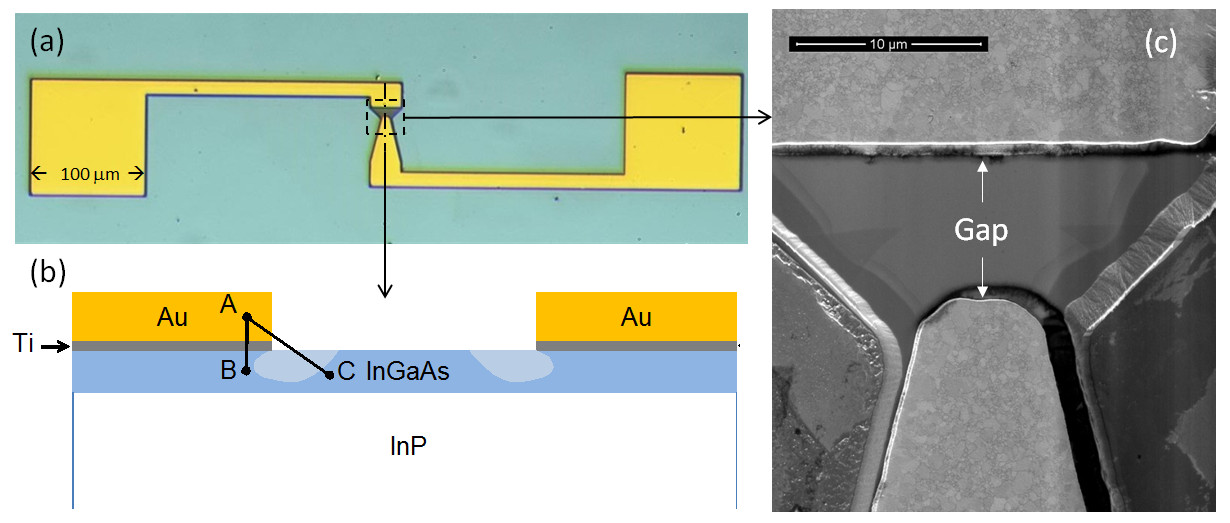}
\caption{(a) Photograph
of a sensor consisting of an active portion with an antenna (central part), bonding pads (large 100$\times$100 areas on the left and right sides) and contact stripes. The dashed box indicates the approximate region corresponding to the cross-sectional sketch shown in (b) and the SEM image shown in (c). (b) Cross-sectional diagram of the modelled device (not to scale). The letters "A", "B", and "C" denote the endpoints of cross sections for which electrostatic analyses will be presented later. The lighter blue areas indicate the approximate positions of the InGaAs regions where the edges of the Au contacts might influence the carrier concentration as a result of electrostatic effects. (c) SEM image of the active region of the sensor, with a gap (as indicated) between the metal contacts.} 
\label{fig:cross}      
\end{figure}

Three InGaAs samples were grown using molecular beam epitaxy on 500~$\upmu$m thick InP:Fe (001) substrates and processed as previously described in \cite{LinasAPL}. The parameters of the investigated samples are summarized in Table~\ref{tab:1}. A cross-sectional diagram of the modelled devices is presented in Fig.~\ref{fig:cross}b. An additional incomplete monolayer (ML) of InAs was incorporated between the InP substrate and the InGaAs layer to improve the crystalline quality of the InGaAs layers during growth in the case of sample II3196. In the other two samples, an InAs interface layer formed between the InP substrate and the active InGaAs layer, with a thickness of approximately 3-5 MLs for sample I197 and 1-2 MLs for sample I204. This layer formed because of the natural desorption of oxides in an arsenic environment, because the SVT MBE system used to grow these samples was not equipped with a phosphorus source. Thus, the sample properties were determined by the InGaAs growth parameters, mainly by the beam equivalent pressure ratio of In to Ga, which influenced the chemical composition of the compounds. Sample I197 contains 47\% In, whereas the In content in sample I204 is 53.2\%. The latter is lattice-matched to the InP substrate.

Mesas were formed via wet chemical etching. For metallization, a Ti layer with a thickness of 20 nm and an Au layer with a thickness of approximately 180 nm were deposited as a two-layer structure. The gap between the metal contacts was set to 10~$\upmu$m during the production of photolithography masks for all samples except I204, for which various gaps of up to 20~$\upmu$m were prepared. The gap sizes are slightly smaller in the finished devices because of processing peculiarities.

\begin{figure}
\centering
\includegraphics[width=1.0\textwidth]{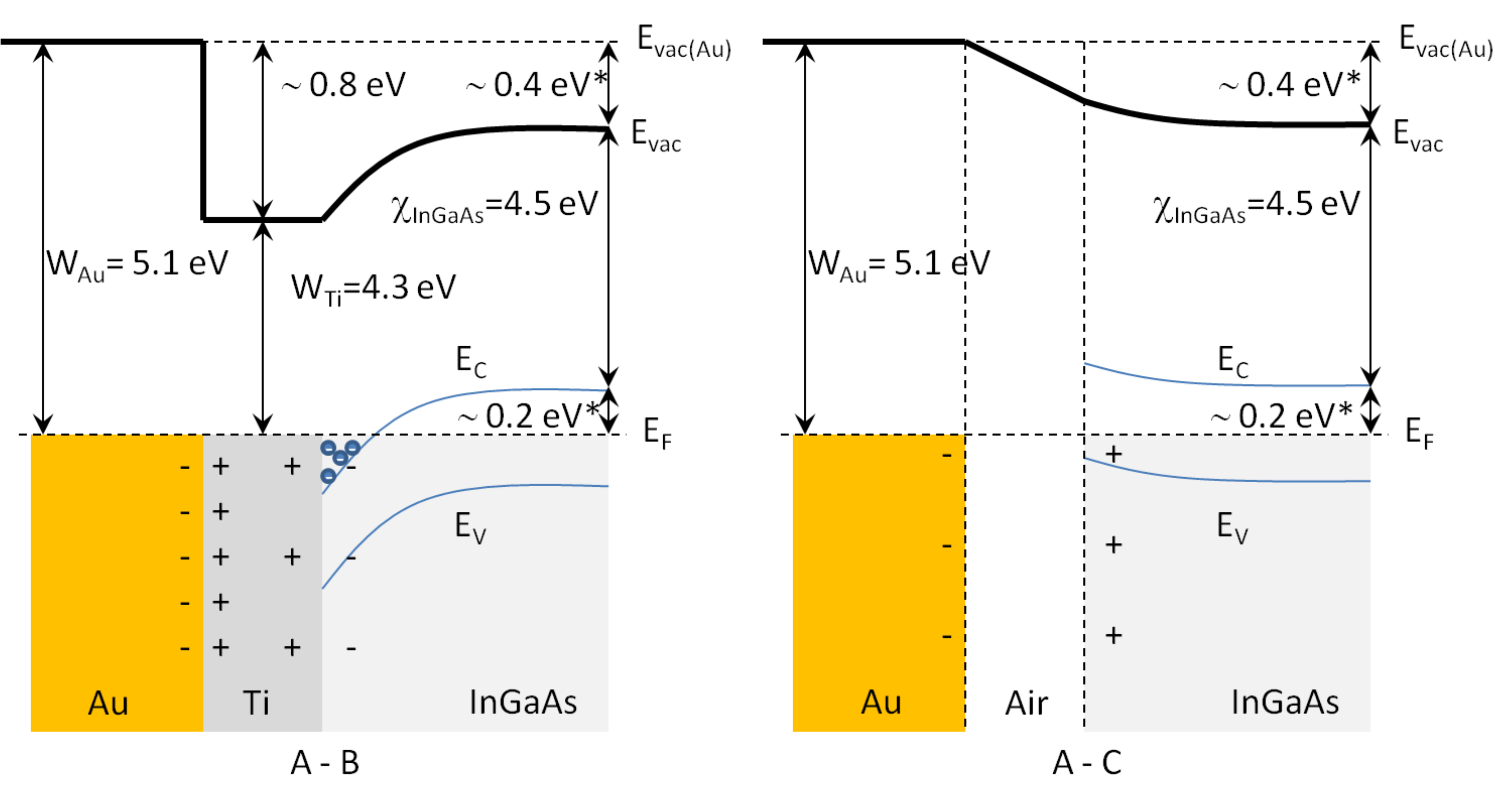}
\caption{Sketches of the expected coordinate dependence of the vacuum energy level (thick line) without the application of an external voltage. Left: along the line from A to B. Right: along the line from A to C. (*) denotes the estimated values for the case of low doping under the assumptions of no surface charge and no Fermi level pinning.}
\label{fig:levels}       
\end{figure}

The scheme of the expected band variation across the tested structures is graphically illustrated in Fig.~\ref{fig:levels}. This figure was drawn under several assumptions: the work functions for gold and titanium are 5.1~eV and 4.3~eV, respectively; the electron affinity is 4.5~eV; and the Fermi level is 0.2~eV below the conduction band minimum. A complete external electric field screening for the bottom of the In\textsubscript{0.53}Ga\textsubscript{0.47}As and only small band bending near the surface of the InGaAs in the vicinity of the gold contacts are assumed. The applicability of these last three assumptions will be discussed later.

Several main features of the energy levels sketched in Fig.~\ref{fig:levels} are worth noting. First, the work function of Ti is smaller than the electron affinity of InGaAs; therefore, a potential well for electrons forms near the interface between these two materials, and a nonalloyed ohmic contact should be formed in the case of $n$-type InGaAs.

It should be noted that the annealing and intermixing of Ti and In atoms at the Ti/InGaAs interface might lead to a reduction in the contact resistance \cite{Lee_stability_98}, but the contact resistance begins to degrade at approximately 400 - 450$^{\circ}$C \cite{Wu1995stability} because of the decomposition of InGaAs. It was also necessary to limit the annealing time to 10~s below 400$^{\circ}$C because of the risk of rapid intermixing of Au and Ti \cite{Martinez_2010} in our simplified metallization stack, which lacked Pt or Pd layers between the Ti and Au.

\begin{table}
\caption{Parameters of the InGaAs samples}
\label{tab:1}       
\begin{tabular}{llllll}
\hline\noalign{\smallskip}
Sample & $t$\textsubscript{InGaAs} (nm) & $(1-x)$(In) & InAs monolayer & Residual doping (cm$^{-3}$)\\
\noalign{\smallskip}\hline\noalign{\smallskip}
II3169 & 520 $\pm$ 10 & 0.53 & $<1$ & $\sim 10^{15}$\\
I197 & 510 $\pm$ 10 & 0.47 & 3-5 & $\sim 10^{16}$  \\
I204 & 525 $\pm$ 10 & 0.532 & 1-2 & $\sim 10^{16}$ \\

\noalign{\smallskip}\hline
\end{tabular}
\end{table}

With no external voltage, the Fermi level is the same along the entire multilayer device;
therefore, a potential difference on the order of several hundred millivolts is expected between the gold layer and the InGaAs layer in regions sufficiently far away from the contact edge for complete screening of the electric field. The situation is more complicated in the case of regions near the edges of the metal contacts, as the side wall of the Au contact layer is located near the InGaAs layer because of the low thickness of the Ti layer (only 20~nm). Therefore, because of the complex shape of this multilayer device, the potential distribution was calculated numerically using the Sentaurus TCAD program. Simulations were performed to numerically compute the possible electrostatic potential distributions for several scenarios involving different overlaps of the metal layers and the surface charges. Experimentally measured current-voltage (IV) curves and the results of Kelvin probe measurements served as a starting point for the initial evaluation of the simulation parameters.

\section{Electrical characterization of the samples}
\label{sec:Elcha}

\subsection{Kelvin probe measurements}
\label{sec:kelvin}

\begin{figure}
\centering
\includegraphics[width=0.7\textwidth]{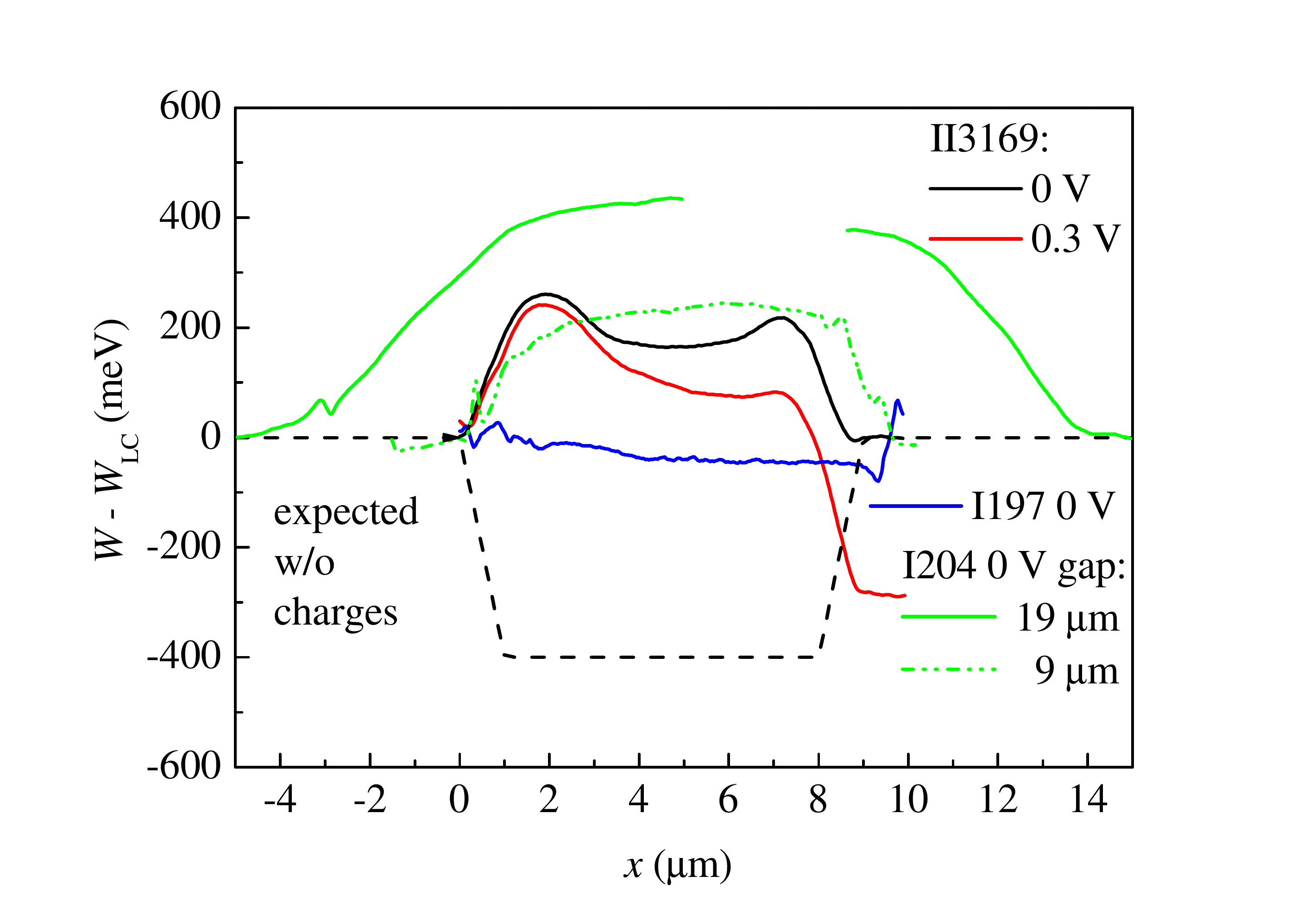}
\caption{Dependence of the work function on the distance between the metal contacts of the BT detectors. Solid lines - KPFM measurement results; dashed line - sketch of the expected dependence (without the influence of surface charges) based on Fig.~\ref{fig:levels}. The work function of the left contact served as the reference value for generating this graph. The edge of the left contact is located at a coordinate of $\it{x}$~≈~0~$\upmu$m for samples II3169, I197, and I204 with 9~$\upmu$m gaps between the contacts and at $\it{x}~≈-$5~$\upmu$m for sample I204 with a 19~$\upmu$m gap. The right edge of the contact is located at $\it{x}$~≈~9~$\upmu$m for samples II3169, I197, and I204 (narrower gap) and at $\it{x}~≈~$14~$\upmu$m for sample I204 (wider gap). Two measured dependences are presented for sample II3169, namely, the dependences with a 0.3~V bias voltage (red line) and without it (black line). Two separate measurements were performed for sample I204 with the wider gap because the gap size (19~$\upmu$m) exceeded the maximum available scanning range in our set-up.}
\label{fig:kelvindist}       
\end{figure}

The work function was measured across the planar structure between the contacts, as shown in Fig.~\ref{fig:cross}b, and the coordinate dependence of the work function was obtained for each tested sample. The measurements were performed using a Veeco Dimension 3100/Nanoscope IVa Atomic Force Microscope. The results obtained for sample II3169 under the application of a 0.3~V external voltage and for the other samples without an external voltage are presented in Fig.~\ref{fig:kelvindist}. Note that the zero coordinate corresponds to the edge of the narrower metal contact. This contact also served as the zero-voltage reference for the measurements. Because of the limitations of the available scanning range in our set-up, it was necessary to perform two scans of the device on sample I204 with a 19~$\upmu$m gap.

As one can clearly see from Fig.~\ref{fig:kelvindist}, the work function difference gradually increases within the first 1~$\upmu$m from the edge of the metal, reaching values of approximately 0.4~eV (sample I204 with a 19~$\upmu$m gap) and >~0.2~eV (sample I204 with a 9~$\upmu$m gap and sample II3169) with respect to the gold surface. Meanwhile, much smaller potential differences are observed for sample I197. A sketch of the expected results based on Fig.~\ref{fig:levels} is also provided in this figure. One can easily observe a difference between the measured and expected results of at least several hundreds of meV in the centre between the contacts. Such a large difference could suggest either that the Fermi level is located several hundred meV lower than initially estimated or that the observed results are influenced by additional factors, such as surface states and/or surface charges. In addition, the work function differences between the devices on sample I204 with 9~$\upmu$m and 19~$\upmu$m gaps offers the idea that the distribution of electrostatic charges within the detector might be influenced by the distance between the contacts.

\subsection{Scanning tunnelling spectroscopy measurements}
\label{sec:stm}

\begin{figure}
\begin{tabular}{cc}
	\includegraphics[width=0.48\textwidth]{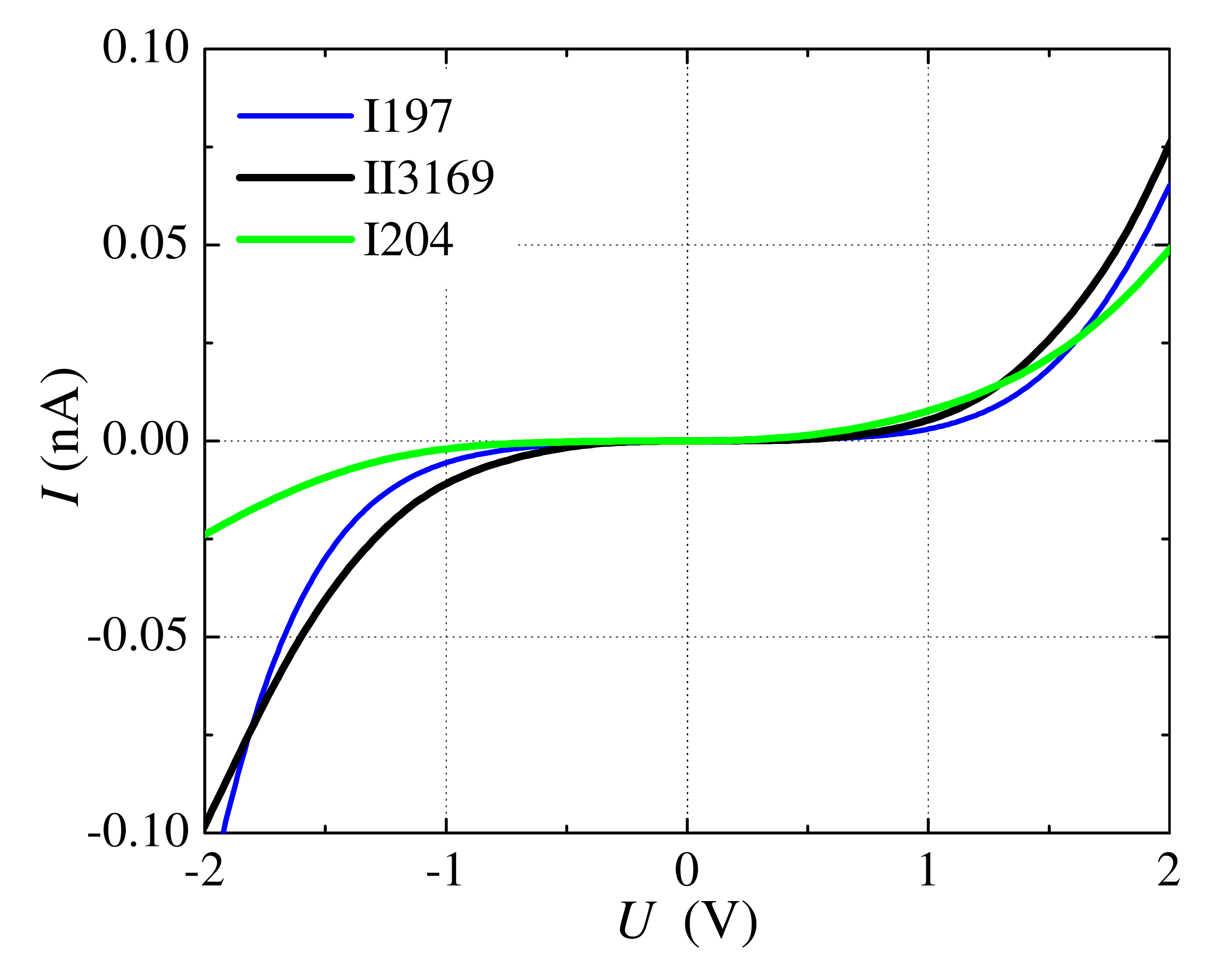} &
	\includegraphics[width=0.48\textwidth]{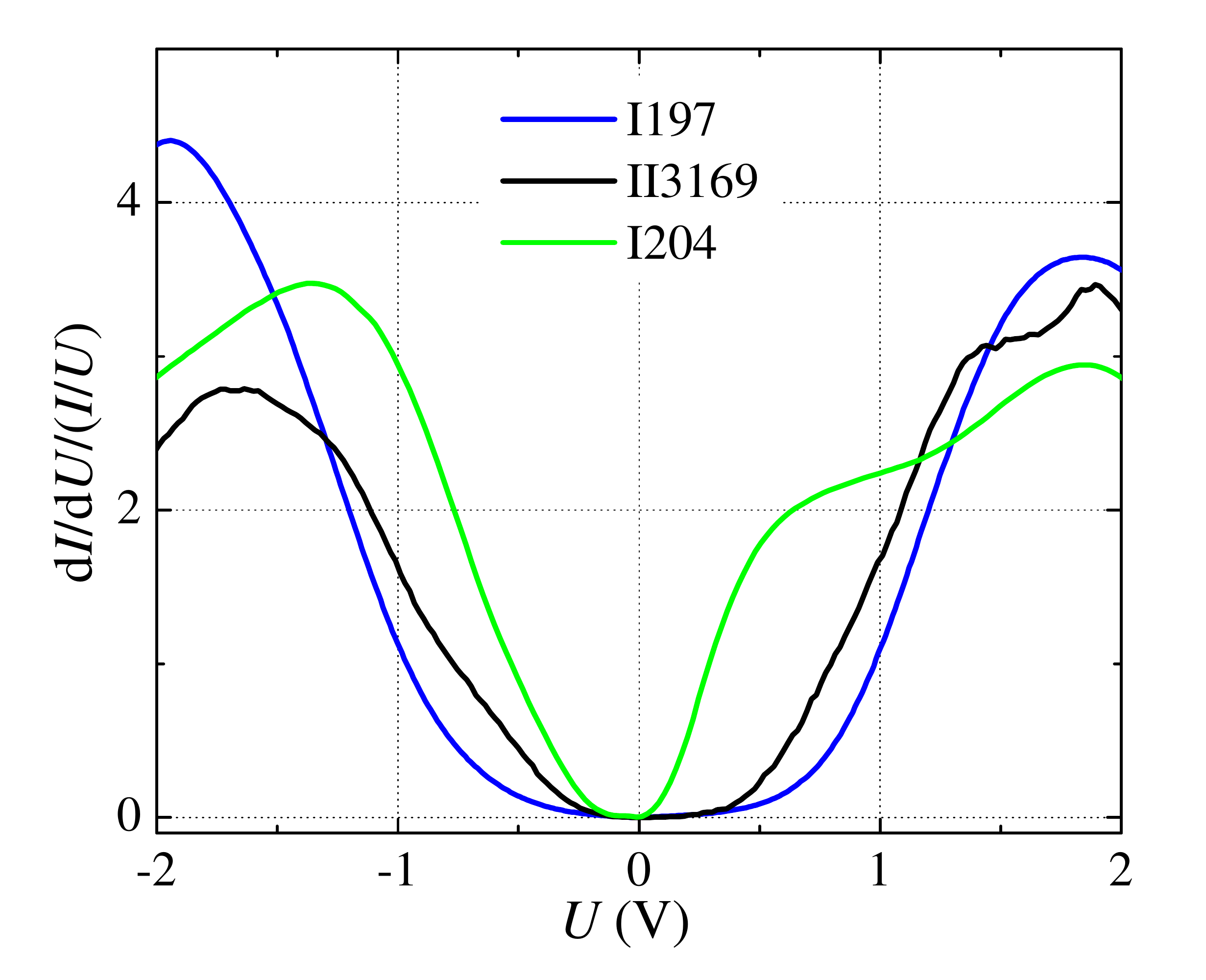} \\
\end{tabular}
\caption{Dependence of the tunnelling current (left) and its derivative (right) on the sample bias.}
\label{fig:stm}       
\end{figure}

The surface Fermi level position and the local density of states (LDOS) were also evaluated via scanning tunnelling spectroscopy (STS) by measuring and analysing the dependence of the local tunnelling current on the applied voltage. The experiments were performed using the same set-up to supplement the previously presented KPFM measurements. The results of these measurements are presented in Fig.~\ref{fig:stm}. As one can see, the normalized derivative of the tunnelling current (and, simultaneously, the surface density of states) begins to increase at approximately 0.4~V and -~0.3~V in the cases of samples I197 and II3196, respectively. The difference between these voltages is consistent with the bandgap of the InGaAs sample. The voltage corresponding to higher LDOS regions indicates that the Fermi level is located only slightly below the middle of the bandgap. Similar {\it{p}}-type behaviour of the surface and pinned Fermi level positions was recently observed by W. Melitz $et~al.$ \cite{melitz2010scanning} in the case of unpassivated {\it{n}}-type InGaAs layers. Fermi level unpinning and more typical {\it{n}}-type behaviour were demonstrated in their work after the deposition of an ordered monolayer of absorbates from trimethylaluminium (TMA) \cite{melitz2010scanning}. 

Unfortunately, reliable conclusions cannot be drawn from this comparison because the production sequence for our devices included exposure to the ambient atmosphere during the photolithography and etching procedures.

The difference of 0.4~eV between the Fermi level and the bottom of the conduction band in combination with a 4.5~eV electron affinity would imply a work function of approximately 4.9~eV, so a reason for the approximately 600~meV difference needs to be found for sample II3169 to explain the coordinate dependence of the work function depicted in Fig.~\ref{fig:kelvindist}. 

A similar difference of several hundred meV between the STS and KPFM results has previously been observed for InGaAs and attributed to a density of charged defects of $7.3-9.3{\times}10^{12}$~cm$^{-2}$, shifting the Fermi level in the case of KPFM and resulting in a much lower influence of surface charges in the case of STS \cite{melitz2010scanning}. 
Since comparable work function differences between the STS and KPFM results are observed for our samples, we employed similar surface charge values in our simulations, as presented later. 

In Fig.~\ref{fig:stm}, the data for sample I204 reveal substantially smaller differences in voltage (and electron energy) between regions with an increased LDOS compared with the two other samples. LDOS differences might lead to different surface charge densities for all of the investigated samples. Together with charge density variations within the InGaAs layers, these differences might lead to the surface potential differences presented in Fig.~\ref{fig:kelvindist}.  

\subsection{Measurements of current-voltage characteristics}
\label{sec:iv}

\begin{figure}
\centering
\includegraphics[width=1.0\textwidth]{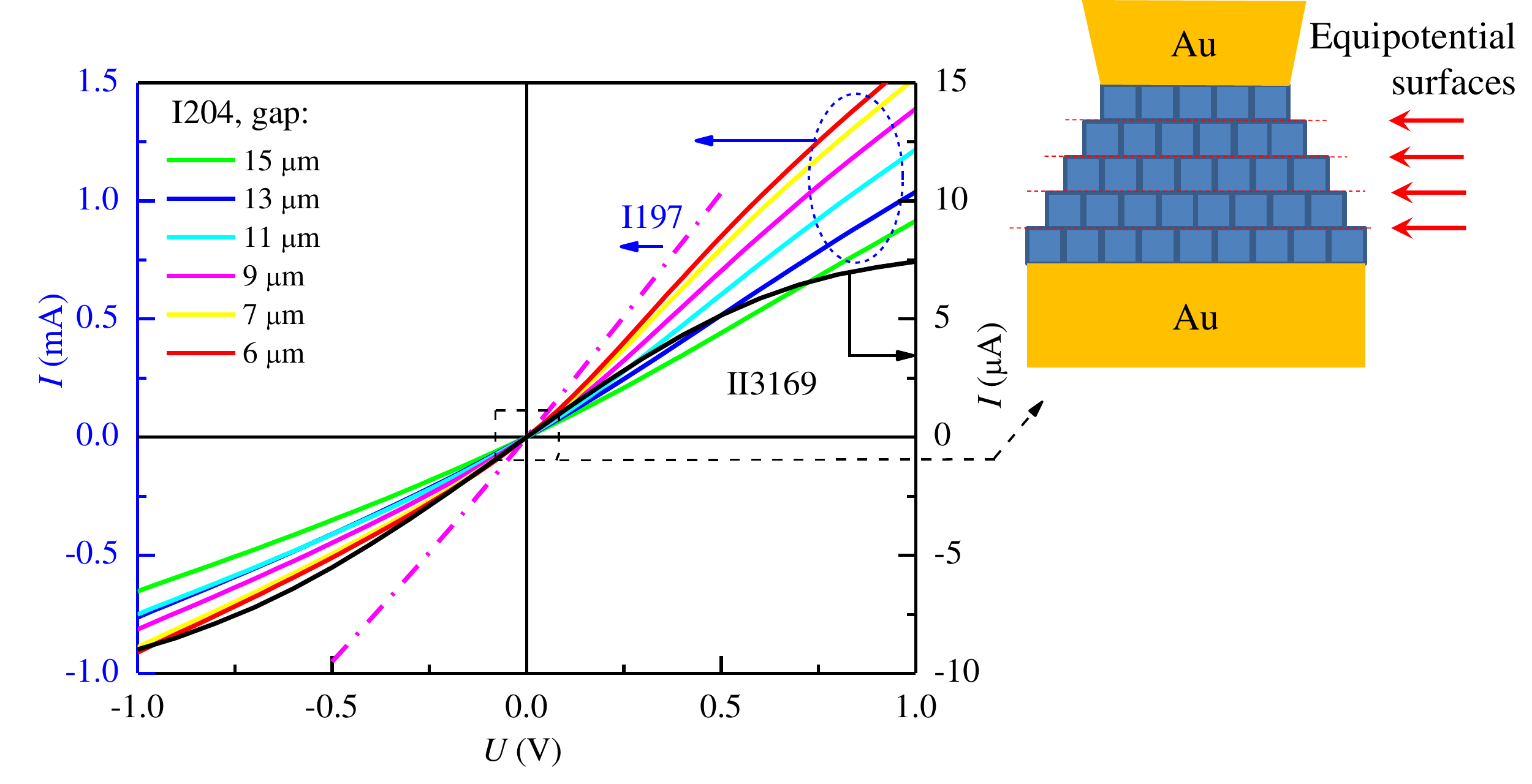}
\caption{Left: IV curves of the studied samples. Note the scale difference. The dotted black rectangle indicates the part of the IV curve that was used to estimate the average carrier concentration. Right: illustration of the model used to relate the differential resistance of the sample to the square resistance of the layer - the area corresponding to the active region of the sensor was filled with small squares, and the resistances were calculated assuming equipotential borders between the rows of squares and parallel connections within these rows.}
\label{fig:iv}       
\end{figure}

The current–voltage (IV) characteristics of the investigated samples are presented in Fig.~\ref{fig:iv}. Two major features are apparent: a difference of nearly two orders of magnitude between the differential resistances at the 0~V point (the slopes of the curves are similar, but the current values on the left and right scales differ by two orders of magnitude) and bending of the IV curve for the sample with the higher resistance as the voltage approaches 1~V. Note that the IV characteristics of sample II3169 are similar to those of a typical BT diode as presented in \cite{KasalynasJAP2011}. A voltage of 1~V, averaged across the approximately 10~$\upmu$m gap between the metal contacts, translates to an average electric field of 1~kV/cm. Considering the shape of the device, one can estimate a maximum electric field value of 2.5~kV/cm near the tip of the narrower metal contact using the simplified model presented in the next paragraph and Fig.~\ref{fig:iv}. This value is of the same order as, but still smaller than, the value of approximately 3~kV/cm at which negative differential mobility is expected to manifest itself \cite{BalynasKrotkusTransport}. Therefore, the IV linearity of sample I204 up to 1~V is not surprising, even with a slight gap variation (Fig.~\ref{fig:iv}). For sample II3169, however, IV curve bending is already observed at 0.5~V; therefore, one must look for additional factors, such as field effects due to surface charges or the edges of the metal contacts, which might lead to additional field non-uniformity within the InGaAs layer. The device based on sample I197 exhibited the highest conductance; therefore, the voltages had to be limited to 0.5~V to avoid damaging the device. Nevertheless, its IV curve is highly linear even near a voltage of 0.5~V, at which the device based on InGaAs sample II3196 already exhibits substantial non-linearity. 

Because of the sample shape, the resistance $R$ of a sample (in $\upOmega$) is expected to be approximately 80~\% of the sheet resistance of the layer, $R$\textsubscript{sq} (in $\upOmega$/sq units). This approximate value can be evaluated by meshing the shape of the active area of the device into rows of square cells. Furthermore, the resistance itself can be calculated, under the assumptions of parallel electrical connections within lines and serial connections between them, as illustrated on the right-hand side of Fig.~\ref{fig:iv}. In this case, the measured differential resistance of sample II3169 is $R~\approx~$100~k$\upOmega$; in combination with an expected electron mobility of $\mu = 9\times10^3$~cm$^2$/V$\cdot$s, this value leads to an average concentration estimate of $n$~=~7$\cdot10^{13}$~cm$^{-3}$ for an \mbox{InGaAs} layer thickness of 0.5~$\upmu$m. This value, in combination with the expected intrinsic carrier density of $n_i=6.3\cdot10^{11} $cm$^{-3}$ \cite{PaulIntrinsic}, leads in turn to a difference of approximately 0.2~eV between the bottom of the InGaAs conduction band and the Fermi level. This value was used to sketch the energy levels farther away from the metal contacts on the right-hand side of Fig.~\ref{fig:levels}. As mentioned earlier, this model also provides an estimate of a ratio of 2.5 between the maximum and average electric field values. However, this value should be treated as an absolute upper limit because of the imperfections of the model used -- it assumes current spreading without potential differences in the lateral direction.

In the fully depleted InGaAs layer approximation and under the assumption of zero electric field strength near the bottom of the InGaAs layer, a concentration of ionized donors of $N_D=n=$7$\cdot10^{13}$cm$^{-3}$, a channel thickness of $t\textsubscript{InGaAs}$~≈~520~nm and a relative dielectric permittivity of $\epsilon = $13.9, one can apply the usual calculation procedure for field-effect transistor (FET) channel pinch-off to estimate the potential difference required for full depletion between the bottom and top of the layer based on Gauss's law:  

\begin{equation}
\upDelta\varphi{\textsubscript{PO}}=\int_0^{t_{\text{InGaAs}}}q_0{\cdot}N_D{\cdot}{\frac{z}{\epsilon\epsilon_0}}{\cdot}{\text{d}}z=q_0{\cdot}N_D{\cdot}{\frac{{t^2_{\text{InGaAs}}}}{2\epsilon\epsilon_0}}~\approx~12~\text{mV},
\label{eq:fi_po}
\end{equation}
where $q_0$ is the elementary charge. The corresponding value of the electron potential energy difference (12 meV) is more than one order of magnitude smaller than the work function differences between the experimentally measured values and the expected ones (see Fig.~\ref{fig:kelvindist}); hence, one can expect that surface charges might play a crucial role in defining the electrical properties of our sensors. At least partial depletion of the layers might also mean that the apparent average carrier concentration is substantially lower that the actual dopant concentration. 

Therefore, Sentaurus TCAD IV simulations of InGaAs layers with the same or higher doping levels and partially depletion by surface charges were performed to reveal the conditions for reduced differential conductance observed at voltages approaching $\pm$1~V (Fig.~\ref{fig:iv}), as observed for sample II3169.

The two orders of magnitude higher conductivities of samples I197 and I204 would translate into higher (also by two orders of magnitude) average carrier concentrations and, consequently, potential differences on the order of volts between the top and bottom of the InGaAs channel for complete pinch-off. There is no evidence of such large differences in Fig.~\ref{fig:kelvindist}; therefore, it is not surprising that these devices exhibit ohmic-like behaviour, unlike sample II3169.

\section{IV simulation results}
\label{sec:ivsim}

IV simulations were performed using Sentaurus TCAD \cite{EZ_synopsys} to reveal possible phenomena related to the observed peculiarities of the IV curves. Since low apparent conductivities in the samples are correlated with current saturation in the IV curves, which is behaviour that indicates IV characteristics similar to those of FETs \cite{Teppe_nano_FET}, two main factors possibly related to device fabrication were considered. 

The first one is overlapping of the Au layers with the Ti sublayers, possibly occurring during the gold deposition process. At points of overlap, the Au layers have tiny interfaces with the InGaAs channel near the Ti contact sublayers. As a result of the larger work function of Au (5~eV), Schottky barriers form around the Ti (4.3~eV) contact sublayers, thereby impeding the movement of electrons from one Ti contact sublayer to another upon the application of voltage.

The second feature that may affect the nature of the IV curves is the presence of surface charge at the top surface of the InGaAs channel between the contacts. A negative surface charge (as expected from Kelvin probe measurements) will create a negative electrostatic potential, thereby forcing the electrons to move away from the InGaAs surface and reducing the thickness of the conductive portion of the channel.

Both of these technological aspects could facilitate the formation of charge depletion regions within the InGaAs channel, which would effectively act like gates in FET devices.

In addition, we paid particular attention to the magnitude of the electron mobility in our simulations, as the electric fields present in the studied devices are sufficiently large to affect charge mobility. Therefore, we separately studied the high-field dependence of the mobility on the shape of the IV curves, as presented later in the paper.

The material parameters used in the device simulations were obtained from the Sentaurus TCAD material database library. The parameter data file for In$_{0.53}$Ga$_{0.47}$As was created for a given molecular fraction as prescribed in \cite{EZ_TCAD_parameters}. In the parameter file, we modified the low-field electron mobility values according to various literature sources. The mobility values used in the simulation are presented in Table \ref{EZ_table_mobility}. Additional simulations at high electric fields were also performed with modified mobility values, as explained in the subsection below.

Based on the results and analysis of our IV measurements, which revealed FET-like behaviour of the II3169 device, it was assumed that the current would be primarily limited by channel pinch-off near one of the contacts; therefore, the distance between the Ti contacts was slightly reduced in comparison with the real devices to conserve computer resources. This distance was fixed at 6~$\upmu$m, whereas other parameters were varied as described in the text. Of course, such an arrangement leads to underestimation of the voltage values in the linear range of the IV curve, but it still permits the evaluation of the conditions that lead to channel pinch-off.

\begin{table}[h!]
  \centering
  \caption{Mobility values used in the simulations}
  \label{EZ_table_mobility}
  \begin{tabular}{l|c|l}
    Doping concentration (cm$^{-3}$) & Mobility (cm$^2$V$^{-1}$s$^{-1}$) & Source\\
    \hline
    $1\cdot 10^{14}$ & 16000 & TCAD database \\
    $9\cdot 10^{14}$ & 10000 & \cite{BalynasKrotkusTransport} \\
    $1\cdot 10^{15}$ & 10000 & same as for $9\cdot 10^{14}$   \\
    $2.9\cdot 10^{15}$ & 8500 & \cite{BalynasKrotkusTransport} \\
    $4\cdot 10^{16}$ & 7000   & \cite{Chattopadhyay1981jpc} \\
    $2.3\cdot 10^{17}$ & 6000 & \cite{Chattopadhyay1981jpc} \\
  \end{tabular}
\end{table}

Physical models used in the simulations were as follows: Fermi statistics at $T=300$~K; hydrodynamic transport for electrons and holes; mobility high-field saturation, either with a hydrodynamic driving force (high-field effect disabled) or with a charge driving force given by the gradient of a quasi-Fermi potential (high-field effect enabled); recombination models including the Shockley-Read-Hall, radiative and Auger models. Schottky barrier models were specified at the Ti/InGaAs and Au/InGaAs interfaces.

In the simulations, two contacts were specified on the top of each Au layer. The boundary conditions were set such that the contact voltages were 0~V on one contact and variable on the other.

The simulated device had the two-dimensional geometry shown in Fig.~\ref{fig:cross}b. The specified layer thicknesses were as follows: InP layer - 1~$\upmu$m, InGaAs layer - 0.536~$\upmu$m, each titanium contact sublayer - 20~nm, each gold contact layer - 100~nm. A 0.1~$\upmu$m air layer on top of the device was used to ensure correct definition of the surface charge at the InGaAs/air interface. The length of the entire simulation region was set to 8~$\upmu$m. 

\subsection{Influence of contact overlaps and surface charges on IV characteristics}
\label{sec:fix_mob}

\begin{figure}
\centering
\includegraphics[width=0.9\textwidth]{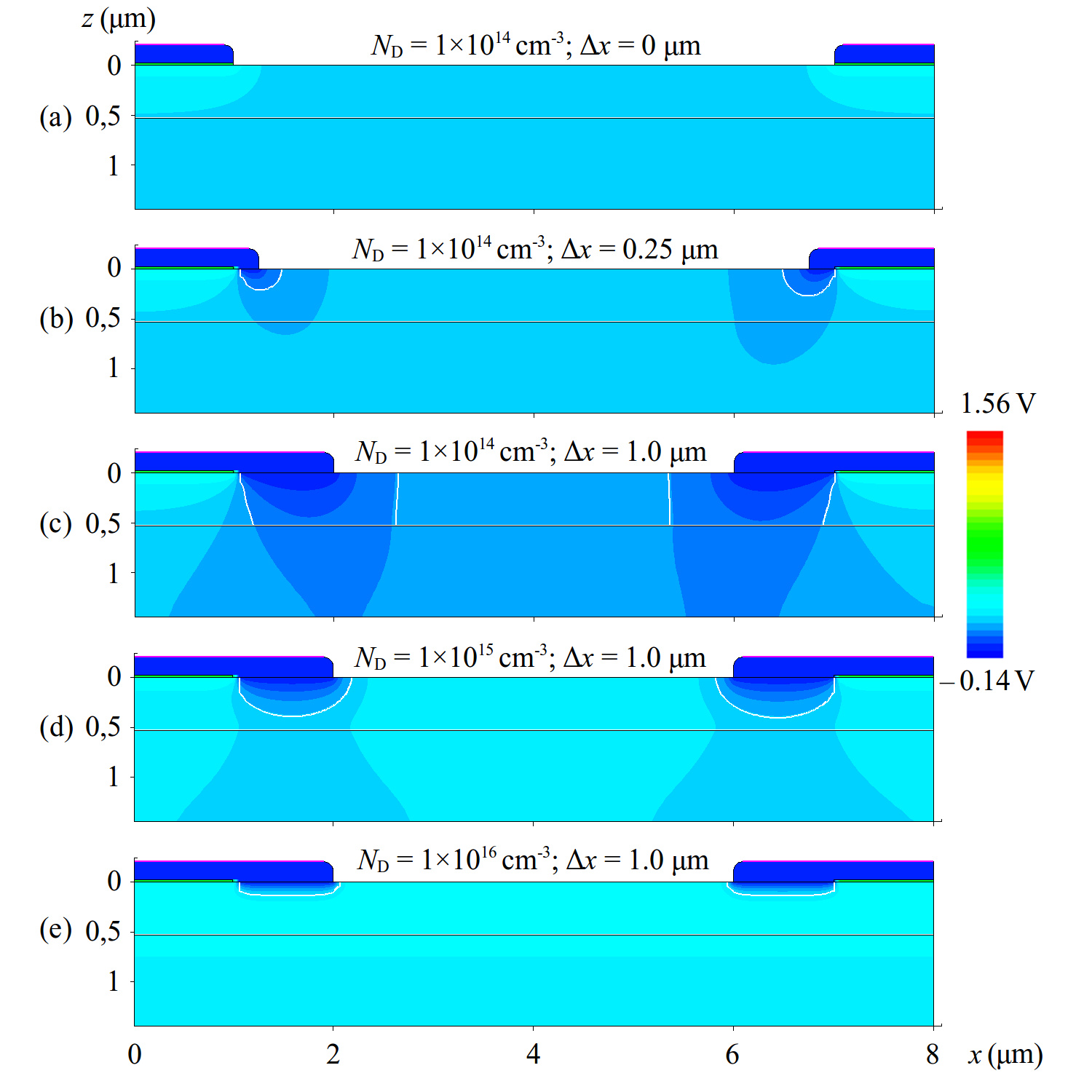}
\caption{Cross-sectional distributions of the electrostatic potential in the device: (a-c) for a fixed $N_D=1.0\cdot10^{14}~$cm$^{-3}$ doping and a variable 0-1.0~$\upmu$m overlap of the Au contacts and InGaAs; (c-e) for a fixed 1.0~$\upmu$m overlap of the Au contacts and InGaAs and a variable $N_D=1\cdot10^{14}$-$1\cdot10^{16}~$cm$^{-3}$ doping.}
\label{fig:overdop0}       
\end{figure}

First, electrostatic (in the absence of an external voltage) simulations were attempted with variable metal overlaps and with variable semiconductor doping levels. The simulation results are summarized in Fig.~\ref{fig:overdop0}. As one can see in panels (a)-(c), at a fixed doping level of $N_D=1.0\cdot10^{14}~$cm$^{-3}$, Schottky barriers and depleted regions (with tentative boundaries as indicated by white lines) form below the gold contacts because of the relatively high work function of gold. The depleted region covers the entire channel once the overlap is increased to 1.0~$\upmu$m. At higher doping levels, as expected, the size of the depleted regions is reduced because of the higher density of charges and, therefore, higher screening of the electrostatic fields by thinner charged layers. 

The results at higher doping levels offer a possible explanation for the experimentally observed IV characteristics (Fig.~\ref{fig:iv}), as it is expected that an open channel and a high carrier concentration will lead to ohmic-like behaviour at low voltages. At lower doping, FET-like current saturation is expected once the channel is closed by the depleted region. However, such a simplified analysis is obstructed by the fact that the InGaAs layer thickness is of the same order as the thickness of the metal and its radius of curvature near the edge; therefore, short-channel-type transistor operation and a full numerical simulation has to be considered. 

\begin{figure}
\centering
\includegraphics[width=0.9\textwidth]{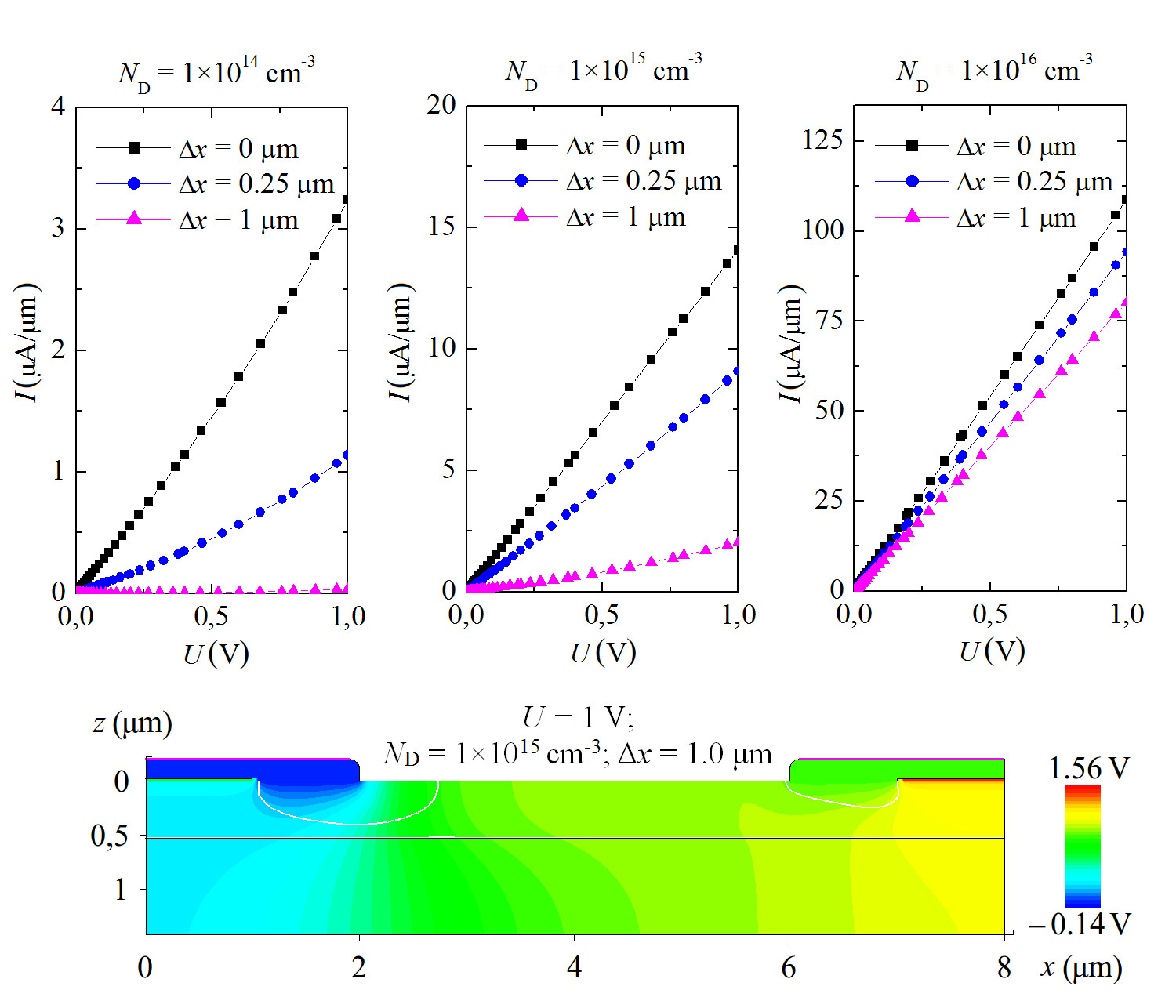}
\caption{Top: simulated IV curves for several doping levels and overlaps of the metal contacts. Bottom: electrostatic potential distribution within the device in the case of a 1~V sample bias, a 1.0~$\upmu$m overlap and a doping level of $N_D=1.0\cdot10^{15}~$cm$^{-3}$.}
\label{fig:overdop1}       
\end{figure}

Results for several metal overlaps and doping concentrations are presented in Fig.~\ref{fig:overdop1}. As one can clearly see, the downward bending observed in Fig.~\ref{fig:iv} is not visible here. Instead, in some cases, the IV curves even bend upwards. These results confirm the previously expressed reservations regarding the formation of Schottky-type gates by means of metal overlaps alone. An additional problem is illustrated in the bottom panel of Fig.~\ref{fig:overdop1}. As the sample bias increases, the thickness of depleted region near one of the metal contacts grows; the channel width is reduced there, and the largest potential drop develops near this contact. This is somewhat in conflict with the results previously presented in Fig.~\ref{fig:kelvindist}, where a rather gradual drop in the additional potential difference is observed.

Both of these factors may be used for further model refinement to consider the possible influence of surface charges on the open InGaAs surface for the simulation of a larger (along the $x$ axis) partially depleted InGaAs region. Several experimental results provide reasonable starting estimates of the surface charge values
and the doping values. Notably, even without a limitation imposed on the current by complete depletion, an average doping level on the order of at least $N_D=1.0\times10^{15}~$cm$^{-3}$~is required (as illustrated in Fig.~\ref{fig:overdop1}) to obtain the experimental current values presented in Fig.~\ref{fig:iv}. 

\begin{figure}
\centering
\includegraphics[width=0.9\textwidth]{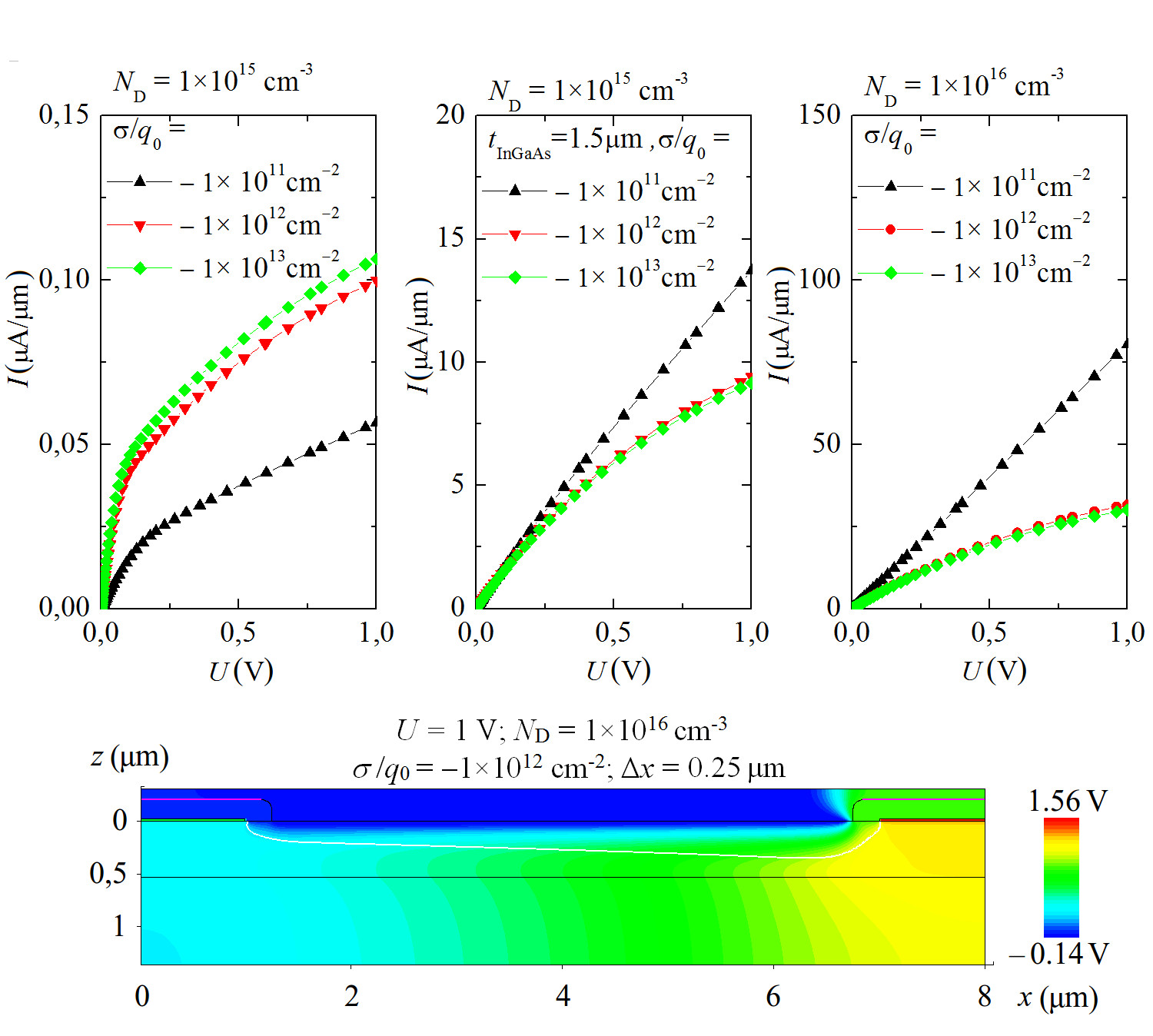}
\caption{Top: simulated IV curves for two doping levels and different surface charge densities. The results in the middle panel were obtained for a larger InGaAs layer thickness of $t$\textsubscript{InGaAs}=~1.5~$\upmu$m. Bottom: electrostatic potential distribution within the device in the case of a 1~V bias, a 0.25~$\upmu$m overlap, a doping level of $N_D=1.0\cdot10^{16}~$cm$^{-3}$ and a surface charge density of $\sigma=-1.0\cdot10^{12}~q_0\times$cm$^{-2}$.}
\label{fig:surfchar}       
\end{figure}

Several simulation results that appear closest to FET-type IV characteristics are presented in Fig.~\ref{fig:surfchar}. As one can clearly see, the introduction of relevant surface charges results in the necessary IV curve bending. 
However, the parameter dependence of the IV curve shape is significantly different in nature, as defined by both the doping level and the width of the InGaAs channel. For a fixed channel width and high doping levels (top right panel in Fig.~\ref{fig:surfchar}), a larger surface charge causes the current value to decrease. This is because the depletion region below the InGaAs/air interface widens with an increasing surface charge density and the conductivity of the channel is effectively decreased. At a doping level of $N_D=1.0\cdot10^{16}~$cm$^{-3}$, the Debye screening length in InGaAs is 0.05~$\upmu$m and the depletion region width is 0.1 - 0.15~$\upmu$m for the given surface charge values. Opposite behaviour is observed in the case of lower doping for the same channel width (top left panel in Fig.~\ref{fig:surfchar}). The current values are two orders of magnitude lower in the presence of surface charge (compare the current values with those without surface charge in Fig.~\ref{fig:overdop1}, top middle panel, blue curve), and the current itself increases with increasing surface charge. Such different behaviour of the IV curves occurs when the depletion region width becomes equal to or larger than the the width of the InGaAs channel. In the case of $N_D=1.0\cdot10^{15}~$cm$^{-3}$~doping, the Debye screening length is 0.14~$\upmu$m and the depletion region width is 0.4 - 0.5~$\upmu$m. In this situation, the role played by the majority carriers (electrons) in the total current is dramatically reduced and a parametric increase in the current occurs because of the increasing minority carrier concentration (recall that the product of the electron and hole concentrations is constant, according to the law of mass action).
However, for the same doping level, the situation changes if the InGaAs channel width is enlarged to be wider than the depletion region. As shown in the top middle panel of Fig.~\ref{fig:surfchar}, the parametric IV dependences for a device with a thicker InGaAs channel are similar for both doping levels.

The magnitude of the obtained current values is lower than that expected from the experimental data recorded at a doping level of $N_D=1.0\cdot10^{15}~$cm$^{-3}$ and higher for $N_D=1.0\cdot10^{16}~$cm$^{-3}$. However, both of these values are higher than the estimated average value (Subsec.~\ref{sec:iv}, $N_D=7.0\cdot10^{13}~$cm$^{-3}$) because of partial channel depletion.

\subsection{Influence of high-field effects}
\label{sec:lim_mob}

\begin{figure}
\centering
\includegraphics[width=1.0\textwidth]{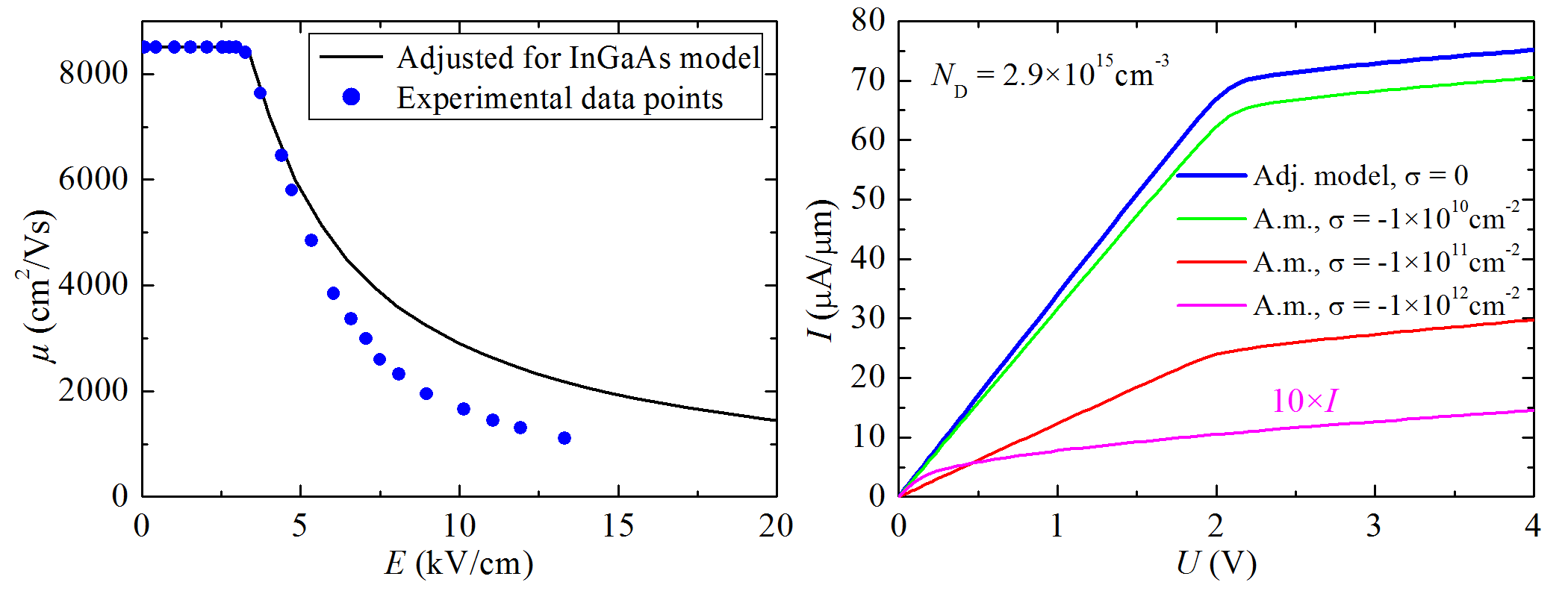}
\caption{(Left) Dependence of the carrier mobility on the electric field used in our calculations and key experimental data points for comparison (after \cite{BalynasKrotkusTransport}). (Right) IV simulation results obtained using an adjusted carrier mobility model based on the model of Canali et al. \cite{canali1975electron} that is built into TCAD, with and without surface charges. }
\label{fig:lim_mob}       
\end{figure}

Since our simulation results revealed substantial carrier density and electrostatic variations in the vertical direction (perpendicular to an unetched InGaAs surface), more comprehensive models were subsequently simulated (Fig.~\ref{fig:lim_mob}). 

The high-field mobility dependence model used in TCAD is usually referred to as the Canali model \cite{canali1975electron}. In fact, it originates from the Caughey–Thomas formula \cite{thomas1967carrier}:
\begin{eqnarray}
\label{EQ_mob-high}
\mu(F) &=& \frac{(\alpha + 1) \mu_0} 
{\alpha + \left[ 1 + \left( \frac{(\alpha + 1) \mu_0 F}{v_{sat}}\right)^\beta \right]^{1/\beta}} \, .
\end{eqnarray}
Here, $\mu_0$ is the low-field mobility; $v_{sat}$ is the electron saturation velocity; $F$ is the driving force for the charge carriers, which is the absolute value of the electric field in the simplest case and the absolute value of the gradient of a quasi-Fermi potential in our calculations; and $\alpha$ and $\beta$ are the parameters that are adjusted to fit the experimental data. The $\mu(E)$ dependence for 
InGaAs was obtained by fitting the experimental data reported in 
\cite{BalynasKrotkusTransport} using formula (\ref{EQ_mob-high}), as shown in Fig. \ref{fig:lim_mob} (left), with the following parameters: $\mu_0=8500~$cm$^2$V$^{-1}$s$^{-1}$, $v_{sat} = 2.9 \cdot 10^7$~cm/s, $\alpha = 0$, and $\beta = 40$.

Modifications were made to the Canali et al. model built into TCAD to specify the dependence of the mobility on the electric field as presented in the left panel of Fig.~\ref{fig:lim_mob}. First, the maximum mobility value was adjusted to match that of InGaAs. The model was further adjusted to ensure a constant mobility up to 3~kV/cm. For comparison, key experimentally obtained points from\cite{BalynasKrotkusTransport} are also presented.

Several simulation results obtained using these mobility models are presented in the right panel of Fig.~\ref{fig:lim_mob}. As one can clearly see, the negative differential resistance at high fields above 3~kV/cm is not sufficient to justify the experimentally observed IV curve shape of the sample with the higher resistance. Linear current growth is restricted to below 1~V only for the case of the largest surface charge, for which the current values are already substantially reduced because of field effects.

\section{FDTD simulations}
\label{sec:fdtd}

\begin{figure}
\centering
\includegraphics[width=0.8\textwidth]{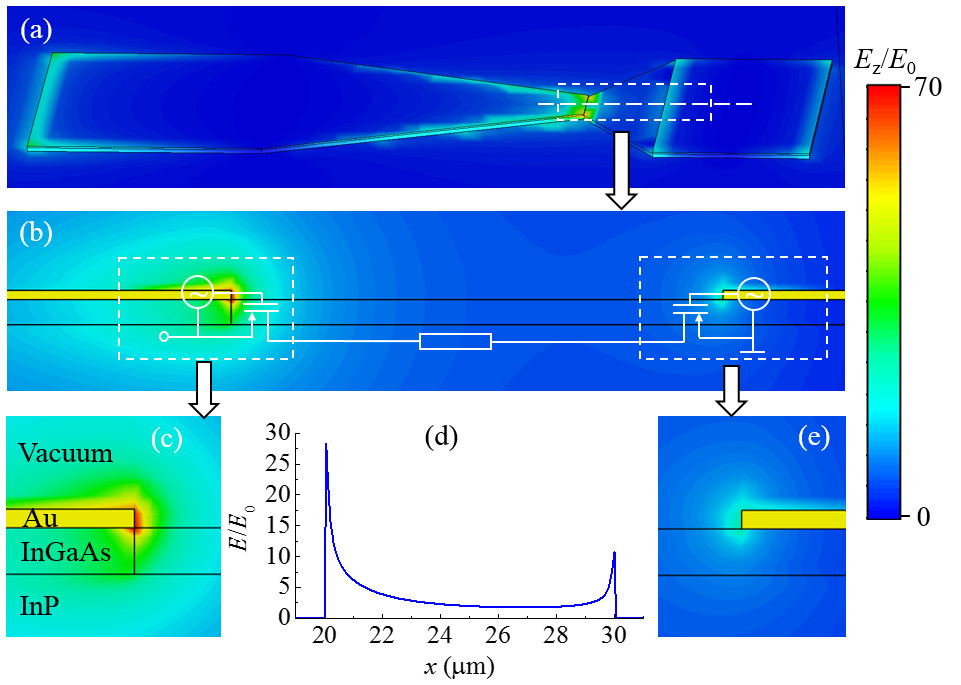}
\caption{Simulated electric field amplitude distribution within the detector at 0.6~THz (a); an enlarged view of part of the top panel and a sketch of the equivalent circuit (b); the electric field amplitude distributions near the edges of the metal contacts (c,e); and the dependence of the amplitude on the coordinate along the symmetry axis of the active region (d). Dashed boxes indicate the approximate locations of the areas enlarged in the subsequent panels of the figure.}
\label{fig:fdtd}       
\end{figure}

Finite-difference time-domain (FDTD) simulations were performed to reveal the expected distributions of the electric field amplitudes near the metal contacts. The obtained electric field amplitude distribution within the InGaAs detector is presented in Fig.~\ref{fig:fdtd}. As one can clearly see, the electric field is mainly concentrated near the edge of the sharper metal contact. This alternating field acts as a high-frequency voltage source between the gate and the source of the equivalent FET. The vertical electric field component $E_y$ is responsible for modulating the channel conductivity. The alternating current induced by the in-plane field component $E_x$ is therefore partially rectified to produce an observable DC component. The second transistor on the other side of the device also acts as a FET detector; however, its influence is reduced because of the much lower electric field amplitudes.
This detection mechanism shows similarities with one that has recently been demonstrated for junctionless FETs \cite{KnapJunctionless}. However, in our case, the edge of the metal contact acts as a compact gate, whereas an equivalent to a DC gate-source voltage would be "provided" by metal work function differences and open surface charges.

\section{Influence of the DC current on the detector sensitivity}
\label{sec:dc_itaka}

Several important detector properties can be expected to follow from the simulation results presented in Sections \ref{sec:ivsim} and \ref{sec:fdtd}. 
First, if the channels of both of the "FET-like" parts of the device are indeed partially conductive in the absence of an applied external voltage (as expected from the IV characteristics shown in Fig.~\ref{fig:iv}) and if a potential reduction of the corresponding metal electrode can selectively "squeeze" one of them (as expected from Fig. ~\ref{fig:overdop1}), then one would expect that it might be possible to tune the resistive self-mixing performance of both "FET-like" parts by changing the applied DC voltage. In addition, both "FET-like" parts are connected in series and should produce rectified voltages of opposite signs. As a result, the voltage of the rectified signal should change sign with the switching of pinch-off from one "FET" to the other. For DC voltages between these two regimes, the detector signal should change gradually and reach zero for equal contributions from both parts of the device.

\subsection{Experimental data}
\label{sec:dc_itaka_exp}

\begin{figure}
\centering
\includegraphics[width=0.80\textwidth]{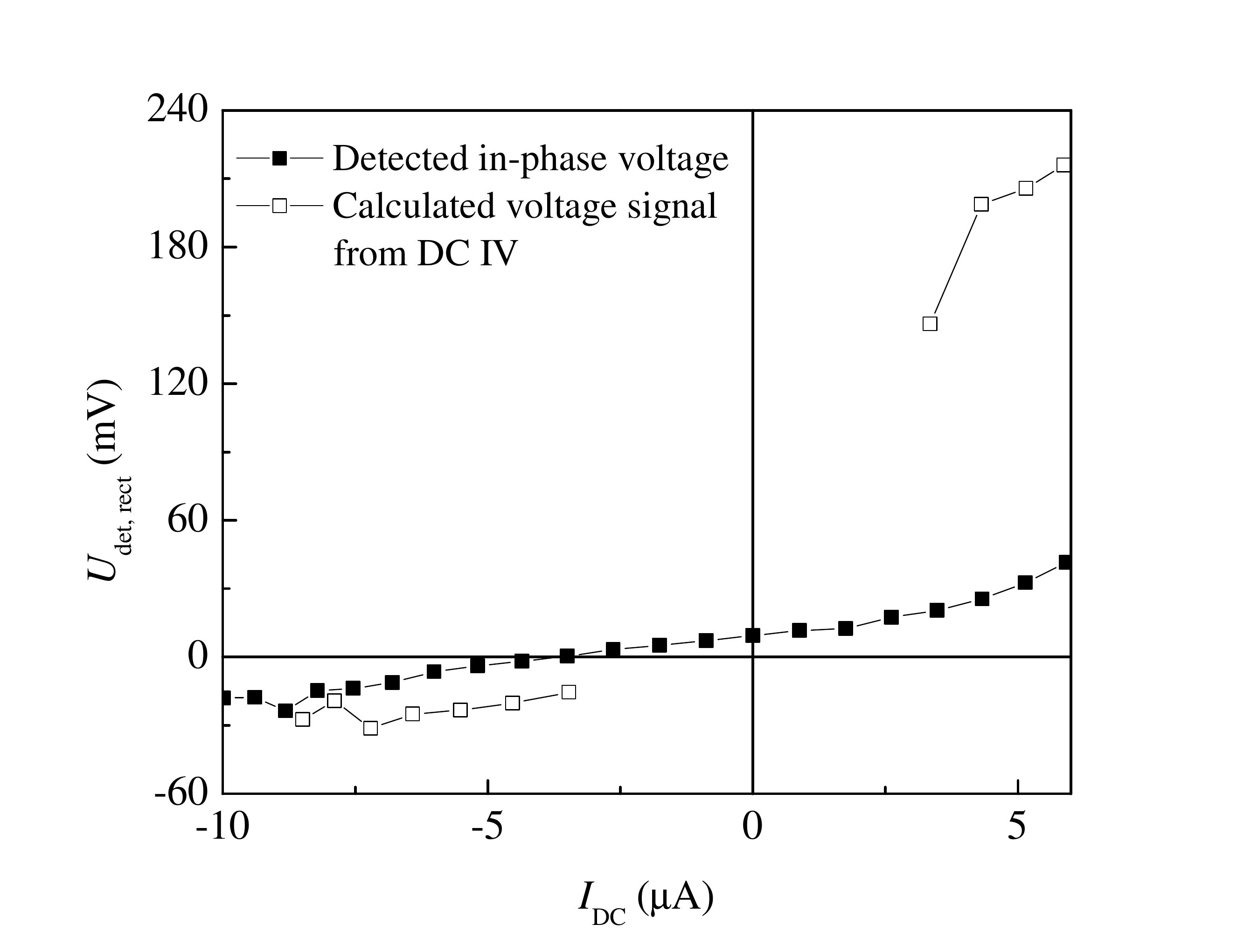}
\caption{Dependence of the calculated and experimentally detected signals on the applied DC current.  A frequency multiplier chain operating at 0.584 THz served as the THz source, with its output modulated at 1 kHz by an external function generator. A Signal Recovery 7265 lock-in amplifier, operating with a 10 s integration time, was used to analyse the output signal.}
\label{fig:dc_inf}       
\end{figure}

To experimentally verify these expectations, one of the BT devices was connected to a programmable Keithley 2400 sourcemeter through serially connected 0.47 M$\upOmega$ resistors. The detector was illuminated by focused 0.584 THz waves from a VDI MC156 frequency multiplier chain fed by an Agilent E8257D synthesizer. The amplitude of the incident THz waves was modulated by TTL-compatible signals from an Agilent 33500B function generator, providing complete on-off modulation. The AC signal from the detector was collected using a Signal Recovery 7265 lock-in amplifier operating with a 10~s integration time. Both the in-phase and out-of-phase signals were recorded using a PC connection to the lock-in amplifier. The obtained results are presented in Fig.~\ref{fig:dc_inf}. As expected, the sign of the in-plane voltage signal changes at approximately -3.5~$\upmu$A, when the increased resistive self-mixing performance of the "wider FET" compensates for the influence of the better electric field concentration (illustrated in Fig.~\ref{fig:fdtd}) within the narrower one.

\subsection{Theoretical evaluation}
\label{sec:dc_itaka_theor}

The electric field amplitude distributions (Fig.~\ref{fig:fdtd}d) together with the IV curves shown in Fig.~\ref{fig:iv} offer an opportunity to evaluate whether the experimentally recorded sensor output signal values exceed the expected values computed from the non-linearity of the IV curves.
 
Supposing that 0.8~mW of power of the THz beam is focused onto a spot of approximately one wavelength ($\approx 500~\upmu$m) in diameter, one can obtain an estimated irradiance of 4~kW/m$^2$ in the focal spot. This translates into an electric field amplitude of $E_0\approx1.7$~kV/m.

All simulated electric field amplitude values in Fig.~\ref{fig:fdtd}d
were normalized to $E_0$ so that these results could be applied to evaluate an equivalent source-gate voltage for each of the FETs in Fig.~\ref{fig:fdtd}b. By integrating the electric field within 2~$\upmu$m from the edge of each metal contact (located at $x_1=20~\upmu$m and $x_2=30~\upmu$m), one can estimate AC voltage amplitudes of approximately $U_{AC1}=15{\times}E_0{\times}1~\upmu$m and $U_{AC2}=6{\times}E_0{\times}1~\upmu$m, or 26~mV and 10~mV, for the narrower and wider contacts, respectively.

If one assumes that the higher non-linearity of the IV curve for sample II3196 near a DC current of $5~\upmu$A is caused by the pinch-off of either transistor channel, it should be possible to evaluate the rectified current $I_{rect}$ from these voltages:

\begin{equation}
I_{rect}=\frac{1}{2}\cdot\frac{\partial^2 I}{\partial U^2}\cdot\frac{U_{AC}^2}{2}.
\label{eq:i_rect}
\end{equation}

The voltage signal that could be obtained on an $R{\approx}100$~k$\upOmega$ resistor would then simply be

\begin{equation}
U_{rect}=R{\cdot}I_{rect}.
\label{eq:i_rect2}
\end{equation}

The results of these calculations are presented in Fig.~\ref{fig:dc_inf}. Values substantially larger than those that were experimentally obtained confirm that the electric field amplitudes concentrated near the apex of the device are more than sufficient to generate the experimentally observed signals in the case of the resistive self-mixing model in the dual-FET equivalent circuit.

\section{Conclusions}
\label{sec:conc}

Kelvin probe force microscopy and scanning tunnelling microscopy measurements of the position dependence of the InGaAs surface work function revealed values that were shifted by up to 0.8~eV from the expected ones. This result is consistent with the influence of a substantial surface charge density. The measured current-voltage (IV) characteristics of InGaAs-based bow-tie detectors revealed a substantial reduction in differential conductivity for the sample with the highest resistivity for voltages exceeding 0.5~V. Our numerical simulations confirmed that similar non-linearity of the IV curve can be obtained as a result of a surface charge density of approximately $-(10^{11}$-$10^{12})\times q_0$ cm$^{-2}$ together with a partial overlap of the Au layer with the InGaAs. Under these conditions, the InGaAs layer will be at least partially depleted and the conductivity of the remaining channel can potentially be influenced by the edges of the metal contacts, which act in a manner equivalent to FET gates. As measured under THz irradiation, the dependence of the sensor output voltage on the DC current confirms that this voltage indeed changes sign, as expected from such a dual-FET model. The results of finite-difference time-domain simulations together with an analysis of the IV curve non-linearity confirmed that the THz-wave-induced voltage amplitudes
are sufficient to produce the output signals expected from such a dual-FET structure with the switching of pinch-off from one transistor to the other. All these findings confirm that previously suggested hot-electron-based operation model of bow-tie detectors must be further updated by accounting for FET-like resistive self-mixing in the case of low-conductivity InGaAs. 

This understanding of such a dual-FET-like device operation model for InGaAs THz detectors opens up possibilities for future device optimization to achieve THz detection with even higher performance. For example, different metallization materials could be employed for the two device contacts to enhance the resistive mixing near one edge and reduce it near the other. In addition, InGaAs surface passivation could potentially be applied to reduce the influence of surface effects, and the possible application of additional conductive layers and contacts that are partially transparent to THz radiation could be studied with the purpose of introducing additional possibilities for InGaAs channel conductivity management.

\begin{acknowledgements}
The authors would like to thank Dr. Klaus~Köhler (Fraunhofer-Institut für Angewandte Festkörperphysik, Freiburg, Germany) for providing InGaAs sample II3196.
\end{acknowledgements}

\bibliographystyle{spphys}        		

\end{document}